\documentclass[a4paper,10pt,twocolumn,bibnotes,aps,superscriptaddress]{revtex4-2}

\usepackage[utf8]{inputenc}
\usepackage{graphicx}
\usepackage{mathtools}
\usepackage{xcolor}
\usepackage{amsmath}
\usepackage{amssymb}
\usepackage{xfrac}
\usepackage{soul}
\usepackage{natbib}

\newcommand\rst{\bgroup\markoverwith{\textcolor{red}{\rule[0.5ex]{2pt}{0.4pt}}}\ULon}

\begin{document}

\title{Peculiar band geometry induced giant shift current in ferroelectric SnTe monolayer}
	
\author{Gan Jin}
\affiliation{Key Laboratory of Quantum Information, University of Science and
	Technology of China, Hefei, Anhui, 230026, People's Republic of China}
	
\author{Lixin He}
	\email[Corresponding author: ]{helx@ustc.edu.cn}
	\affiliation{Key Laboratory of Quantum Information, University of Science and
		Technology of China, Hefei, Anhui, 230026, People's Republic of China}
	\affiliation{Institute of Artificial Intelligence, Hefei Comprehensive National Science Center,
		Hefei, Anhui, 230026, People's Republic of China}

\begin{abstract}	
The bulk photovoltaic effect (BPVE) refers to the phenomenon of generating photocurrent or photovoltage in homogeneous noncentrosymmetric materials under illumination, and the intrinsic contribution to the BPVE is known as the shift current effect. We calculate the shift current conductivities of the ferroelectric SnTe monolayer using first-principles methods. We find that the monolayer SnTe has giant shift-current conductivity near the valley points. More remarkably,  the linear optical absorption coefficient at this energy is very small, and therefore leads to an enormous Glass coefficient that is four orders of magnitude larger than that of BaTiO$_3$. The unusual shift-current effects are further investigated using a three-band model. We find that the giant shift current conductivities and Glass coefficient are induced by the nontrivial energy band geometries near the valley points, where the shift-vector diverges. This is a prominent example that the band geometry can play essential roles in the fundamental properties of solids.
\end{abstract}

\maketitle

\section*{Introduction}

The study of BPVE has a long history~\cite{Sturman_1992, Baltz_1981, Sipe_2000}, and recently it has attracted great renewed interest because it potentially allows
the energy conversion efficiency to surpass the  Shockley–Queisser limit~\cite{Shockley_1961, Spanier_2016}.
The shift-current effects are believed to be the main intrinsic contribution to the BPVE~\cite{Sipe_2000, Nastos_2010},
which can be used as an alternative to the photocurrent generated by traditional semiconductor p–n junctions \cite{Tan_2016, Cook_2017}.
It has been demonstrated that the photovoltage generated by shift-current effects can be far above the band gap~\cite{Glass_1974, Dalba_1995, Yang_2010, Alexe_2011}.
		
The high priority of current studies in the field is to find novel materials that have high shift-current conductivities.
Cook et al. proposed design principles of the shift current materials through an effective two-dimensional model and successfully
applied them to monochalcogenide GeS \cite{Cook_2017}.
In addition to conventional ferroelectric materials~\cite{Bhatnagar_2013, Ogawa_2017},
special attention has been given to Weyl semimetals because of the unique topological nature of band structures.
Osterhoudt et al. discovered a huge mid-infrared BPVE in the Weyl semimetal TaAs, which is linked to the topological properties of the material \cite{Osterhoudt_2019}.
Type-II Weyl semimetal TaIrTe$_4$ has been found to have a huge optical response, and the shift current is related
to the divergent Berry curvature at the Weyl nodes \cite{Ma_2019}.
Ahn et al. theoretically  studied the low-frequency properties of BPVEs in topological semimetals, and revealed the relation between the shift current as well as the injection current and the quantum geometry of the material near the Weyl point \cite{Ahn_2020}.
		
In this work, we investigate the nonlinear optical properties of the two-dimensional ferroelectric material SnTe monolayer~\cite{Chang_2016}
using first-principles methods. We find that it has giant shift current conductivities near the valley points.
More remarkably,  the linear optical absorption coefficient at this energy is very small, which leads to an enormous Glass coefficient
of four orders of magnitude larger than that of bulk BaTiO$_3$~\cite{Sturman_1992,Zenkevich_2014, Spanier_2016}.
We develop a minimal three-band model to analyze the mechanism of the giant shift-current effect in the SnTe monolayer.
We find that the giant shift-current effects are induced by the nontrivial band structure geometry, where the shift-vector diverges at
the valley point. We further show that the giant shift-current is related to the
derivatives of the imaginary part of the quantum geometric tensor near the point.
The mechanism is different from the previous works on the Weyl semimetals~\cite{Ahn_2020},
and therefore opens a new play ground for the fascinating physical properties that are determined by the band structure geometries.

\section*{Results}

\subsection*{Crystal structure}

\begin{figure}[htbp]
				\centering
				\includegraphics[width=0.5\textwidth]{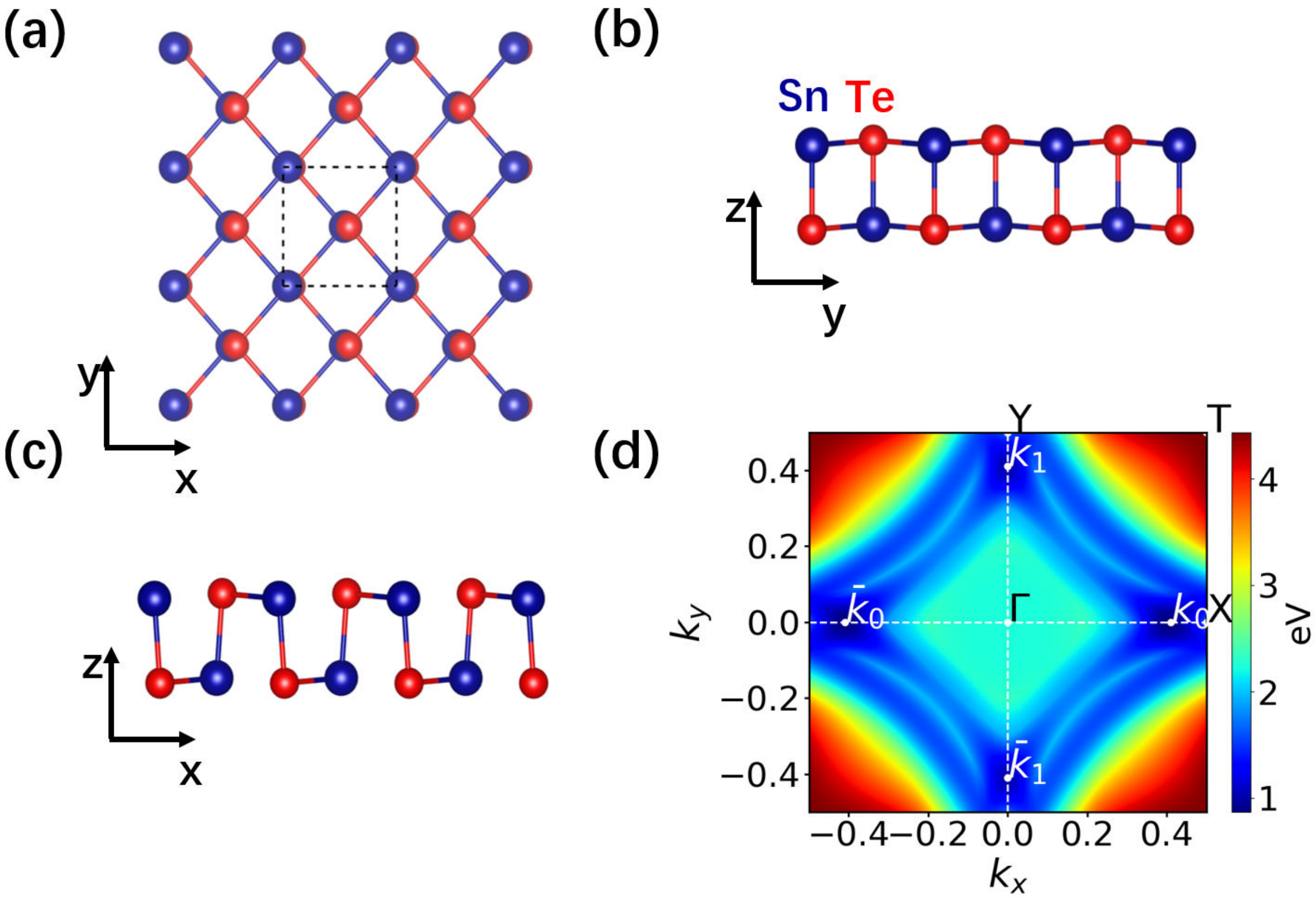}
				\caption{(a) Top view, (b) side view along the $x$ direction and (c) side view along the $y$ direction of the crystal structure of the SnTe monolayer. Sn and Te atoms are shown using blue and red spheres, respectively. The unit cell is indicated by the dashed rectangle in (a). (d) The band gap between the lowest conduction band and the highest valence band in the Brillouin zone, where  $\mathbf{k}_0$, $\bar{\mathbf{k}}_0$ $\mathbf{k}_1$ and $\bar{\mathbf{k}}_1$ are the valley points.
				}
				\label{fig:structure}
			\end{figure}

In 2016, Chang et al. \cite{Chang_2016} discovered that the SnTe monolayer has robust in-plane ferroelectricity
with a Curie temperature as high as 270 K, which is greatly enhanced from its bulk value of 98 K.
As a member of the Group IV monochalcogenide
(MX, M = Ge, Sn, Pb; X = S, Se, Te) family,
the SnTe monolayer, which has great potential in miniaturized ferroelectric devices, has been extensively studied
experimentally \cite{Chang_2019} and via first-principles calculations \cite{Wan_2017, Absor_2019, Kim_2019}.
The structure of the SnTe monolayer is shown in Fig. \ref{fig:structure}, which has a hinge-like structure similar to that of phosphorene.
The SnTe monolayer has a Pmn2$_1$ space group with mirror symmetry ($M_{xz}$) and glide mirror symmetry ($G$).
It has an in-plane ferroelectricity along the $x$-axis \cite{Wan_2017, Chang_2019, Absor_2019, Kim_2019}.
			
The lack of inversion symmetry suggests that  the SnTe monolayer should have shift-current effects.
We perform first-principles calculations to investigate the shift-current effects in the SnTe monolayer.
Details of the calculations are presented in the METHODS section.
			
\subsection*{Band structure and shift current conductivities}

The band structures of SnTe monolayer with and without spin-orbit coupling (SOC) are shown in Fig.S1 of the Supplementary Information (SI).
There are four valley points of the band structure in the first Brillouin zone.  Two valley points, $\mathbf{k}_0$=(0.41, 0.0), $\mathbf{k}_1$=(0.0, 0.41), in direct coordinates, are on the $\Gamma$-$X$ line and $\Gamma$-$Y$ line respectively, whereas the other two valley points are obtained by the time inversion
symmetry of the above two $k$-points, i.e., $\bar{\mathbf{k}}_0=-\mathbf{k}_0$ and $\bar{\mathbf{k}}_1=-\mathbf{k}_1$,
as shown in Fig. \ref{fig:structure}(d).
The introduction of SOC only slightly changes the positions of the valley points (see Fig. S1 in SI).
			
The break of inversion symmetry would lead to the shift current in the SnTe monolayer,
i.e., a nonlinear dc photocurrent under illumination~\cite{Sturman_1992, Baltz_1981, Sipe_2000},
\begin{equation}
J^{a} = 2 \sigma^{abc}(0; \omega, -\omega) E_{b}(\omega) E_{c}(-\omega),
\end{equation}
where $\sigma^{abc}(0; \omega, -\omega)$ is the shift-current conductivity, and $a$ ($b$, $c$)=$x$, $y$, $z$ are the crystal axes.
Due to the 2D monolayer structure, we do not consider the current in the $z$-direction.
Similarly, we do not consider the electric field applied in the $z$-direction.
Because of the mirror symmetry $M_{xz}$ of the SnTe monolayer,
only $\sigma^{xxx}$, $\sigma^{xyy}$, and $\sigma^{yxy}$=$\sigma^{yyx}$ are nonzero (see the SI).
			
We investigate the shift-current conductivities by using first-principles calculations \cite{Sipe_2000, Young_2012, Azpiroz_2018, Wang_2019},
 and the numerical results are consistent with the above symmetry analysis.
The 3D-like conductivities are obtained assuming an active single-layer thickness of 3.12 $\text{\AA}$\cite{Rangel_2017, Azpiroz_2018}.
Although quantitatively, the results with and without SOC are somehow different, but the inclusion of SOC  does not affect the main results and conclusions of the paper (see Fig. S4 in SI). Therefore, we only discuss the results without SOC here, and the analysis can be equally applied to the results with SOC.
			
Figure 2(a) depicts $\sigma^{yxy}$, which is the largest component of the shift-current conductivities.
$\sigma^{yxy}$ has three distinct peaks at 0.87 eV, 1.24 eV, and 2.18 eV, respectively, in the energy interval 0$\sim$4 eV.
The highest peak is at $\hbar \omega_0$=0.87 eV, with  $\sigma^{yxy}_{3D}$=481.89 $\mu$A/V$^2$, which is significantly
larger than the known high BPVE material GeS of the order of 150 $\mu$A/V$^2$ and 250 $\mu$A/V$^2$
in state-of-the-art Si-based solar cells \cite{Rangel_2017, Azpiroz_2018, Pagliaro_2008}.
			
To experimentally explore this effect, we may apply light with an electric field polarized along the [110] direction, and measure the shift current along the $y$ direction,
which gives
\begin{equation}
j^{y}=2 \sigma^{yxy} E_x(\omega) E_y(-\omega).
\end{equation}

In previous works \cite{Tan_2016, Cook_2017, Azpiroz_2018}, it has been suggested that to have a large shift current, a large joint density of states (JDOS) is necessary,
which has been used as a design principle in finding materials with a large shift current.
Surprisingly, we find that the JDOS at $\hbar \omega_0$=0.87 eV is extremely small as plotted in Fig. S2 in the SI.
As a consequence, the absorption coefficients are expected to be small at this photon energy.
Indeed, the absorption coefficient $\alpha^{[110]}$ is also very small around $\hbar\omega_0$=0.87 eV, as shown in Fig. 2(b).
What is more amazing is that the absorption coefficient $\alpha^{yy}$=0 at this energy, strongly against our intuition, given that $\sigma^{yxy}$ is huge.
			
\begin{figure}[htbp]
\centering
\includegraphics[width=0.5\textwidth]{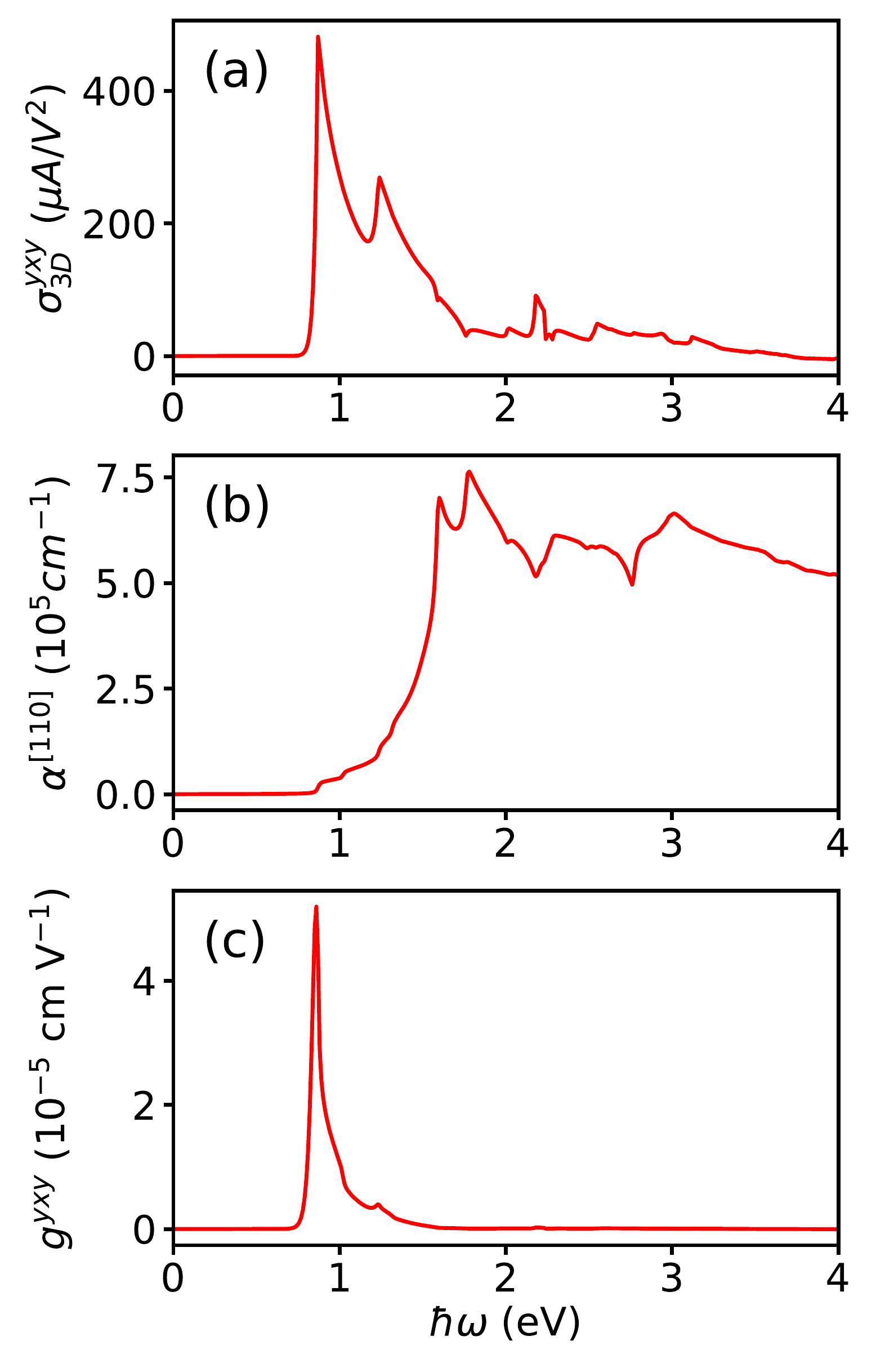}
\caption{(a) Shift current conductivity $\sigma^{yxy}_{3D}$ of the SnTe monolayer. There are three peaks at 0.87, 1.24, and 2.18 eV, respectively. The highest peak is at 0.87 eV. (b) The absorption coefficient $\alpha^{[110]}$ and (c) the Glass coefficient of the SnTe monolayer, respectively. The Glass coefficient has a sharp peak at 0.87 eV.}
\label{fig:shift}
\end{figure}

The above results have important physical consequences.
We compute the Glass coefficient~\cite{Glass_1974, Sturman_1992, Young_2012, Tan_2016}
\begin{equation}
g^{abc} = \alpha^{-1} \sigma^{abc} ,
\end{equation}
and the result of $g^{yxy}$ is shown in Fig. \ref{fig:shift}(c).
$g^{yxy}$ has a sharp peak at $\hbar \omega_0$=0.87 eV, due to the giant $\sigma^{yxy}$ and small $\alpha^{[110]}$.
The calculated Glass coefficient $g^{yxy}$ of the SnTe monolayer at $\hbar \omega_0$=0.87 eV is $5.2\times10^{-5}$ cm$\cdot$V$^{-1}$ \cite{Rangel_2017},
which is four orders of magnitude higher than $g_{31}$=3$\times10^{-9}$ cm$\cdot$V$^{-1}$
of the bulk (001)-oriented BaTiO$_3$ crystal \cite{Sturman_1992,Zenkevich_2014, Spanier_2016}.
The Glass coefficient $g$ plays essential roles in the shift current related physical properties.
For example, the photovoltaic field generated by the shift current \cite{Sturman_1992,Spanier_2016} can be estimated as,
\begin{equation}
E_{\mathrm{pv}} \approx \frac{g}{\phi(\mu \tau)_{\mathrm{pv}}} \frac{\hbar \omega}{e} \, ,
\end{equation}
where $\phi$ is the quantum yield, $\hbar\omega$ is the incident photon energy and $\mu$ and $\tau$ are the
mobility and lifetime of the carriers responsible for photoconductivity.
A very large $g$ will lead to a very large photovoltaic field $E_{\mathrm{pv}}$, as in this case, the ``leaking'' current
due to photoconductivity is much weaker than the shift current.
The photovoltaic power conversion efficiency is also closely related to the Glass coefficient~\cite{Sturman_1992}.
			
\section*{Discussion}
	
\subsection*{Origin of the giant shift current in $yxy$ direction}

It is particularly interesting and important to explore the underlaying mechanism of the giant shift-current conductivity and Glass coefficient in the SnTe monolayer. The shift-current tensor is given by \cite{Sipe_2000},
\begin{equation}
\label{eq:shift_current}
\begin{aligned}
\sigma^{abc}(0 ; \omega,-\omega) =	&\frac{\pi e^3}{\hbar^2} \int \frac{d \mathbf{k}}{8 \pi^3} \sum_{n, m} f_{nm} \operatorname{Im}\left[ I_{mn}^{abc} \right.\\
&\left. + I_{mn}^{acb} \right] \delta\left(\omega_{m n}-\omega\right) \, ,
\end{aligned}
\end{equation}
where  $f_{nm}$=$f_n-f_m$ and $\hbar \omega_{nm}$=$E_m-E_n$ are differences
between Fermi occupation factors and band energies, respectively.
$I_{mn}^{abc} = r_{mn}^{b} r_{nm;a}^{c}$, where $r_{nm}^a$ is the inter-band dipole matrix, and
$r_{nm; a}^b$ is the generalized derivative of the dipole matrix, i.e.,
\begin{eqnarray}
r_{nm}^{a}  &=& (1-\delta_{nm}) A_{nm}^{a}, \\
r_{nm;b}^{a} &=& \partial_{b} r_{nm}^{a} - i\left(A_{nn}^{b} - A_{mm}^{b}\right) r_{nm}^{a} .
\end{eqnarray}
Here, $A_{nm}^{a}$ is the non-Abelian Berry connection.
More detailed calculations of $r_{nm}^{a}$ and $r_{nm;b}^{a}$ are described in Methods.
Although $r_{nm}^{a}$ and $r_{nm; a}^b$ are gauge dependent, their norm $|r_{nm}^{a}|$ and $|r_{nm; a}^b|$, as well as $I_{mn}^{abc}$
are gauge invariant~\cite{Azpiroz_2018}. Note that $r_{nm}^{a}$ and $r_{nm; a}^b$ and $I_{mn}^{abc}$ are all ${\bf k}$ dependent, but here we drop the $k$ index for simplicity.
			
We calculate $\sum_{n,m}f_{nm}\operatorname{Im}\left[I_{mn}^{yxy}+I_{mn}^{yyx}\right]\delta\left(\omega_{mn}-\omega_0\right)$ ,
for $\hbar \omega_0$=0.87 eV, in the first BZ [see Fig. S3(a) of SI].
We find the contribution solely comes from the transition between the highest valence band and the lowest conduction band, around the valley points $\mathbf{k}_0$ and ${\mathbf{\bar k}}_0$. Because only the valley points contribute to the optical transitions at $\hbar \omega_0$,
the corresponding JDOS and linear absorption coefficient is very small.
Furthermore, as shown in Fig. S3(b) of SI, $I_{vc}=\operatorname{Im}\left[I_{vc}^{yxy} + I_{vc}^{yyx}\right]$, where $v$ and $c$, are the highest valence band and the lowest conduction band, has a sharp peak at $\mathbf{k}_0$, which leads to the giant shift-current conductivity.
			
\begin{figure}[htbp]
\centering
\includegraphics[width=0.5\textwidth]{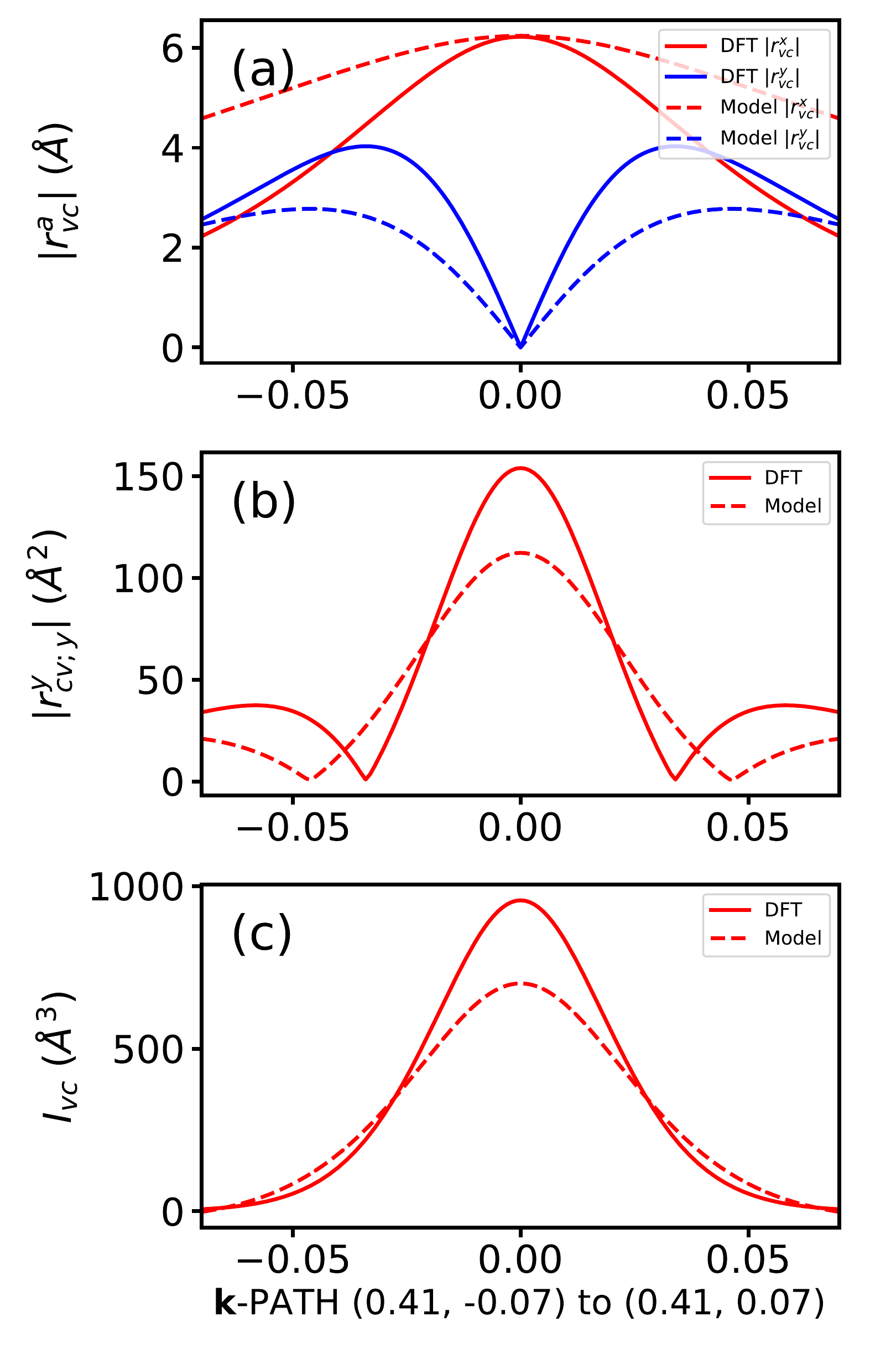}
\caption{
(a) The  norm of dipole matrix $r_{vc}^{x}$ and $r_{vc}^{y}$ along $k_y$, where $v$ and $c$ are highest valence band and lowest conduction band, respectively.
(b) The norm of the generalized derivative of dipole matrix, $r_{cv;y}^{y}$, along  $k_y$. (d) The $I_{vc}$=$\operatorname{Im}\left[I_{vc}^{yxy} + I_{vc}^{yyx}\right]$ along $k_y$. The results obtained from DFT and model calculations are shown in solid and dashed lines, respectively. }
\label{fig:Inm}
\end{figure}

Figure \ref{fig:Inm}(a) depicts the norm of $r_{vc}^x$  and $r_{vc}^y$ along $k_y$ passing through $\mathbf{k_0}$,
which are shown in red and blue solid lines respectively.
We see $|r_{vc}^x|$ has a maximum at $k_y$=0, whereas $|r_{vc}^y|$=0 at $\mathbf{k_0}$ due to the mirror symmetry $M_{xz}$.
However, as seen from Fig. \ref{fig:Inm}(a), $|r_{vc}^y|$ changes rapidly along $k_y$ around $\mathbf{k}_0$.
One may speculate that $r_{vc}^y$  may have a large partial derivative (Eq. 7) along $k_y$ at the valley point.
Indeed, as shown in Fig. \ref{fig:Inm}(b), $|r_{cv;y}^{y}|$ has a peak at $\mathbf{k}_0$.
We plot $I_{vc}$ in Fig. \ref{fig:Inm}(c). Interestingly, we find that $I_{vc} \approx |r_{vc}^{x}||r_{vc;y}^{y}|$ around $\mathbf{k}_0$
(See SI).
Both $r_{vc}^x$ and $r_{cv;y}^{y}$ reach the maximum at ${\bf k}_0$, which lead to the giant  $I_{vc}$.
			
\subsection*{Three-band model}
			
To gain a deeper insight into the physics, we construct a minimal three-band model around the valley point $\mathbf{k}_0$,
\begin{equation}
H(\mathbf{k}) = H_{0} + \mathbb{A} \delta k_{x} + \mathbb{B} \delta k_{y} + \mathbb{C} \delta k_{x}^2 + \mathbb{D} \delta k_{x} \delta k_{y} + \mathbb{E} \delta k_{y}^2,
\end{equation}
up to the quadratic terms of $\delta \mathbf{k}$ , where $\delta \mathbf{k}$=$\mathbf{k}-\mathbf{k}_0$.
More specifically, $\delta k_x$=$k_x-0.41$, $\delta k_y$=$k_y$. We therefore do not distinguish $k_y$ and $\delta k_y$ below.
In the model, the 1st-band is the highest valence band, whereas the 2nd-band and  3rd-band correspond to the lowest
two conduction bands in the DFT calculations.
We fit the Hamiltonian matrix $H(\mathbf{k})$ and velocity matrix $v_{mn}(\mathbf{k})$ from DFT calculations around $\mathbf{k}_0$. The fitted parameters of the model are given in the SI.
			
We first (numerically) calculate $|r_{12}^{x}|$, $|r_{12;y}^{y}|$, and $I_{12}$ of the three-band model, and the results are
compared with those of DFT calculations in Fig. \ref{fig:Inm}(a),(b), and (c), respectively.
The three-band model can semi-quantitatively reproduce the results of DFT calculations, and the discrepancies are due to neglecting other bands.
			
Around $\mathbf{k}_0$, the wave functions of the three-band model can be solved analytically via a second order perturbation theory, and therefore $r_{nm}^{a}$ and $r_{nm;b}^{a}$ can also be calculated analytically.
Especially, at $\mathbf{k}_0$ we have
$v_{nm}^{x}(\mathbf{k}_0) = \mathbb{A}_{nm}$ and $v_{nm}^{y}(\mathbf{k}_0) = \mathbb{B}_{nm}$, therefore
\begin{equation}
r_{nm}^{x}(\mathbf{k_0}) = \frac{\mathbb{A}_{nm}}{i\omega_{nm}}, \quad r_{nm}^{y}(\mathbf{k_0}) = \frac{\mathbb{B}_{nm}}{i\omega_{nm}}.
\end{equation}
Because $\mathbb{A}_{12}$$\neq$0, and  $\mathbb{B}_{12}$=0,  which is actually imposed by the mirror symmetry,
we have $r_{12}^{x}(\mathbf{k}_0) \neq 0$, but $r_{12}^{y}(\mathbf{k}_0) = 0$. This means that the linear (or direct) optical
transition between the 1st and 2nd bands along $k_y$ is forbidden.
Similarly we calculate the $r_{nm;a}^{b}$ of the valley point $\mathbf{k}_0$ [see Eq. (S14) in SI],
and we have,
\begin{equation}
\begin{aligned}
			r_{21;y}^{y}(\mathbf{k_0}) &= \frac{i}{\omega_{21}}\left[\mathbb{B}_{23} \mathbb{B}_{31}\left(\frac{1}{\omega_{31}} + \frac{1}{\omega_{32}}\right) -2\mathbb{E}_{21}\right] .
\end{aligned}
\end{equation}
In the model, the value of $r_{21,y}^{y}$ (and therefore  $I_{12}^{yxy}$)  depends mainly
on the virtual transitions, $\mathbb{B}_{23} \mathbb{B}_{31}$, which corresponding to the last term in Eq. \ref{eq:drnm}.
The first term in Eq.~\ref{eq:drnm} vanishes, because $\mathbb{B}_{12}$=0.
This is remarkable that the linear (direct) optical transition between the 1st-band and the 2nd-band in the $y$ direction is forbidden, but the nonlinear transition may occur because both the 1st-band and 2nd-band have strong coupling with the 3rd-band, which leads to the giant shift-current effect. This effect is quite different from that of the two-band models for Weyl semimetals~\cite{Ahn_2020}.
			
The giant shift-current effects have even more profound origins.
An alternative expression for the shift-current conductivity is written as~\cite{Ahn_2020},
\begin{equation}
\begin{aligned}
\sigma^{yxy}=-\frac{\pi e^3}{\hbar^2} \int_{\mathbf{k}} &\sum_{n, m} f_{nm} \left(R_{mn;y}^{y}-R_{nm;y}^{x}\right) r_{nm}^{x} r_{mn}^{y} \\
& \delta\left(\omega_{mn}-\omega\right) \, ,
\end{aligned}
\end{equation}
where,
\begin{equation}\label{eq:shift-vector}
R_{mn;a}^{b} = i\partial_{a} \ln r_{mn}^{b} + A_{mm}^{a} - A_{nn}^{a},
\end{equation}
is known as the shift vector, which characterizes the displacement of electrons
in real space during the inter-band transition \cite{Sturman_1992, Sturman_2020, Jiang_2021}.
The shift vector is a gauge invariant quantity and can be viewed as a quantum
geometric potential~\cite{Wu_2008}.
According to the perturbation theory, near ${\bf k}_0$,
$r_{12}^{y}(\mathbf{k}) = f_4 \, k_y +f_5\, \delta k_x k_y$, where $f_4$ and $f_5$ are constants. In the model,  $A_{11}^{a}(\mathbf{k}_0) = 0$ and $A_{22}^{a}(\mathbf{k}_0) = 0$  at ${\bf k}_0$, and both are small around
$\mathbf{k}_0$. We can therefore neglect them in the following calculations.
As $\mathbf{k}$ approaches $\mathbf{k}_0$ along $k_y$, i.e., $\delta k_x$=0, we have,
\begin{equation}
R_{12;y}^{y}=i\partial_{y}\ln \left( f_4 \,  k_y \right) ={i\over k_y} \, ,
\end{equation}
i.e., $R_{12;y}^{y}$ is purely imaginary and goes to infinity.
Therefore, ${\bf k}_0$ is a singular point for the shift vector $R_{12;y}^{y}$, which is a monopole in $k$-space.
When $k_y$ is approaching zero, $r_{12}^{y}$ is also approaching zero as discussed in previous sections, and
$R_{21;y}^{x}r_{21}^{x}r_{12}^{y}$ vanishes, but
$R_{12;y}^{y}r_{21}^{x}r_{12}^{y}$ is still finite (actually very large) and purely real (see Fig. 3).
The shift vectors also play important roles in second harmonic generation~\cite{Ahn_2020, Morimoto_2016, Nagaosa_2017}.
It is therefore expected that the SnTe monolayer would has non-trivial second harmonic responses.
The divergent of the shift vector at the ``optical zero'' (i.e., $r_{cv} = 0$) was discussed in Ref.~\cite{Fregoso_2017}. However, the relation between the giant shift current and the divergent shift vector is not revealed.
			
Very recently, nonlinear optical transitions have been related to the Riemannian geometry of the energy bands \cite{Ahn_2020, Ahn_2022}.
We may define the quantum geometric tensor between two bands $m$ and $n$,
\begin{equation}
\label{eq:quantum_metric_tensor}
			Q_{b a}^{mn}=r_{nm}^b r_{mn}^a \equiv g_{ba}^{mn}-\frac{i}{2} F_{ba}^{mn} \, , \, \, a, b=x,y,z
\end{equation}
where $g_{ba}^{mn}$ is the band-resolved quantum metric, $F_{ba}^{mn}$ is the band-resolved Berry curvature, and the two geometric quantities are related to U(1) quantum metric and Berry curvature as $g_{ba}^{n} = \sum_{m\neq n} g_{ba}^{mn}$ and $\Omega_{c}^{n} = \sum_{m\neq n}\epsilon_{cba}F_{ba}^{mn}/2$~\cite{Gao_2020, Watanabe_2021}.

\begin{figure}[tbp]
\centering
\includegraphics[width=0.5\textwidth]{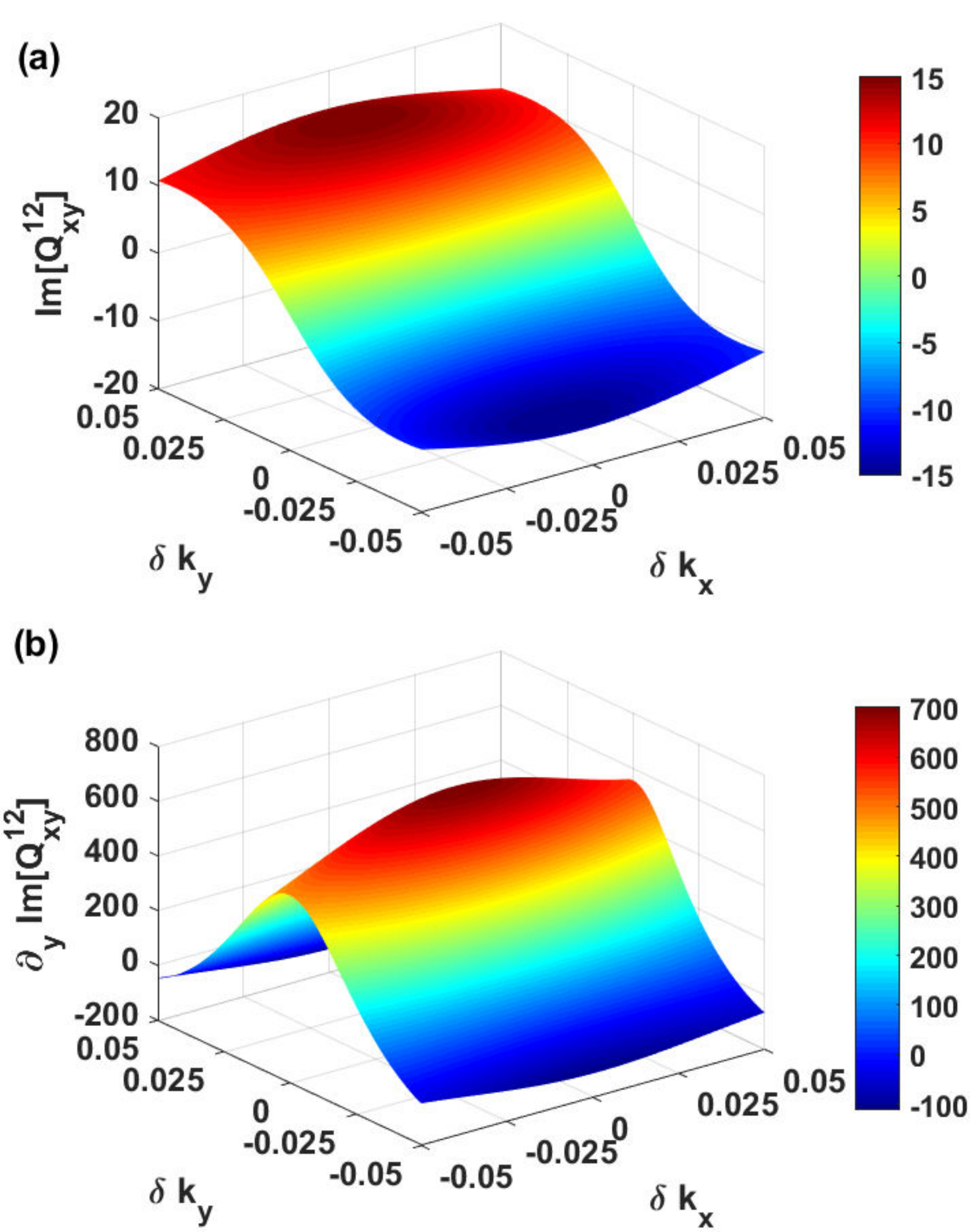}
\caption{The distributions of (a) $\operatorname{Im}\left[Q_{xy}^{12} \right] $ and (b) $\partial_{y} \operatorname{Im}\left[Q_{xy}^{12}\right]$
around the valley point $\mathbf{k}_0$. }
\label{fig:Qxy}
\end{figure}

In our case, we consider the transition between the 1st band and the 2nd band,  and the quantum geometric tensor of the two bands
is given by $Q_{xy}^{12} = r_{12}^{x} r_{21}^{y}$.
We have $Q_{xy}^{12} (\mathbf{k_0}) = 0$ because $r_{21}^{y}(\mathbf{k_0}) = 0$.
However, we show that the partial derivative of the imaginary part of  $Q_{xy}^{12} $
is related to $I_{12}^{yxy}$, i.e.,
\begin{equation}
\operatorname{Im} \left[I_{12}^{yxy}(\mathbf{k_0})\right] =  \partial_{y} \operatorname{Im}  \left.\left[Q_{xy}^{12}
\right]\right|_{\mathbf{k} = \mathbf{k}_0} = -\frac{1}{2} \left.\partial_{y} F_{xy}^{12} \right|_{\mathbf{k} = \mathbf{k}_0}.
\end{equation}
Figure \ref{fig:Qxy}(a) illustrates the distribution of $\operatorname{Im}[Q_{xy}^{12}]$ near $\mathbf{k}_0$ and its derivative with respect to $k_y$
is shown in Fig. \ref{fig:Qxy}(b).
$\operatorname{Im}[Q_{xy}^{12}]$ has a maximum (minimum) at $\delta k_y $=0.05 ($\delta k_y$=-0.05) and $\delta k_x$=0,
which is very similar to the Berry curvature distribution in Fig. 2a of Ref. \cite{Kim_2019}.
Furthermore, $\operatorname{Im}[Q_{xy}^{12}]$ changes rapidly around $\delta k_y$=0, and as a consequence,
$\operatorname{Im}\left[I_{12}^{yxy}(\mathbf{k}_0)\right]$ has a maximum at $\mathbf{k}_0$, which is consistent with that from direct calculations.
			
We have shown that the giant shift-current and Glass coefficient are directly induced by the nontrivial geometry of the energy band near the valley points.
There is no reason that the diverging  shift vector is unique to
the SnTe monolayer. Experimentally, the giant Glass coefficient is a good sign for the diverging shift-vector. However, the phenomena are best observed when the singular points are at the band edge, where they are isolated.
If the singular points are in the middle of the energy bands, the signal may be covered by the light absorption from other $k$ points.
			
\subsection*{CONCLUSIONS}
			
We find a huge shift current effect as well as an enormously large Glass coefficient in the  ferroelectric SnTe monolayer.
These unusual effects are induced by the non-trivial energy band geometry near the valley points, where the shift-vector diverges.
This is another eminent example in which the geometry of the Bloch state plays a profound role in the fundamental properties of a solid.
Whereas most previous examples focus on the ground state properties of the solids, this example shows the case of excitation
in terms of nonlinear optical transitions, which may have great potential applications in photoelectric devices.

\section*{Methods}
\label{sec:method}

The first-principles calculations are carried out with the Atomic orbital Based Ab-initio Computation at UStc (ABACUS) code \cite{Chen_mohan2010,Lipengfei2016}
within the Perdew–Burke–Ernzerhof generalized gradient approximation (GGA) for the exchange–correlation functional \cite{Perdew1996}.
The ABACUS code is developed to perform large-scale density functional theory calculations based
on numerical atomic orbitals (NAO)~\cite{Chen_mohan2010, Lin_2021}.
The optimized norm-conserving Vanderbilt (ONCV) \cite{Hamann2013}
fully relativistic  pseudopotentials \cite{Theurich2001} from the PseudoDojo library\cite{Van2018} are used.
The valence electrons for Sn, Te are $4d^{10} 5s^2 5p^2$, and $4d^{10} 5s^2 5p^4$, and the NAO bases for Sn and Te are 2s2p2d1f and 2s2p2d1f, respectively~\cite{Lin_2021}.
		
In the self-consistent and band structure calculations, the energy cut-off for the wave functions is set to 150 Ry.
The Brillouin zone is sampled using a $\Gamma$-centered 16$\times$16$\times$1 $k$-point mesh.
The structure is fully optimized until all forces are less than 1 meV/\AA.
		
After the self-consistent calculations, the tight-binding Hamiltonian,
\begin{equation}
H_{\mu\nu}({\mathbf{R}})= \langle \mathbf{0}\mu | H | \mathbf{R}\nu\rangle ,
\label{eq:HR}
\end{equation}
the overlap matrices,
\begin{equation}
S_{\mu\nu}({\mathbf{R}})= \langle \mathbf{0}\mu | \mathbf{R}\nu\rangle ,
\label{eq:SR}
\end{equation}
and the dipole matrices (between the NAOs),
\begin{equation}
\mathbf{r}_{\mu\nu}({\mathbf{R}})= \langle \mathbf{0}\mu | \mathbf{r} | \mathbf{R}\nu\rangle ,
\label{eq:rR}
\end{equation}
in the NAO bases are generated, where
$|\mathbf{R}\nu\rangle$ =$\phi_{\nu}\left(\mathbf{r}-\tau_{\nu}-\mathbf{R}\right)$ is the $\nu$-th NAO in the $\mathbf{R}$-th cell,
and $\tau_{\nu}$ is the center of the $\nu$-th NAO  in the unit cell.

The dipole matrix $r_{nm}^{a}$ and its generalized derivative $r_{nm;b}^{a}$  in the shift current Eq.~(\ref{eq:shift_current})
are calculated as follows~\cite{Azpiroz_2018, Cook_2017, Sipe_2000},
\begin{equation}\label{eq:rnm}
r_{nm}^{a} = \frac{v_{nm}^{a}}{i\omega_{nm}} \quad (m \neq n) ,
\end{equation}
and
\begin{equation}\label{eq:drnm}
\begin{aligned}
r_{nm;b}^{a}=& \frac{i}{\omega_{nm}}\left[\frac{v_{nm}^a \Delta_{nm}^b+v_{nm}^b \Delta_{nm}^a}{\omega_{nm}}-w_{nm}^{a b}\right.\\
&\left.+\sum_{p \neq n,m}\left(\frac{v_{np}^a v_{pm}^b}{\omega_{pm}}-\frac{v_{np}^b v_{pm}^a}{\omega_{np}}\right)\right] \quad (m \neq n),
\end{aligned}
\end{equation}
where,
\begin{equation}
\begin{aligned}
v_{n m}^a & = \frac{1}{\hbar}\left\langle u_{n\mathbf{k}}\left|\partial_a H(\mathbf{k})\right| u_{m\mathbf{k}}\right\rangle , \\
\Delta_{n m}^a & =\partial_a \omega_{n m}=v_{n n}^a - v_{m m}^a, \\
w_{n m}^{a b} & =\frac{1}{\hbar}\left\langle u_{n\mathbf{k}}\left|\partial_{a b}^2 H(\mathbf{k})\right| u_{m\mathbf{k}}\right\rangle.
\end{aligned}
\end{equation}
The velocity matrix elements $v_{nm}^a$ are calculated by the {\it ab initio} tight-binding Hamiltonian Eqs. (\ref{eq:HR}) - (\ref{eq:rR})~\cite{Lee_2018,jin_2021}.

The band structures and the optical properties, such as the shift current are calculated using the tight-binding Hamiltonian implemented
in the PY-ATB code~\cite{jin_2023}.

\section*{Data availability}

All data generated and/or analysed during this study are included in this article.

\section*{Code availability}

The ABACUS code is an open source DFT code under the GPL 3.0 licence, which is available from http://abacus.ustc.edu.cn.
The Py-ATB code, also under the GPL 3.0 licence, can be downloaded from https://github.com/jingan-181/pyatb.

\begin{acknowledgments}
This work was funded by the Chinese National Science Foundation Grant Number 12134012. The numerical calculations were performed on the USTC HPC facilities.
\end{acknowledgments}

\section*{Author contributions}

L. He  conducted the project. G. Jin developed the computer code and performed the calculations under the supervision of L. He.
Both authors analyzed the results and wrote the manuscript.

\section*{Competing interests}

The authors declare no competing interests.


%

\end{document}